\documentclass[twocolumn]{aastex63}
\pdfoutput=1 %for arXiv submission
\usepackage{amsmath}
\usepackage{mathtools}
\usepackage{apjfonts}
\usepackage{multirow}

\newcommand{\HarvardPhysics}{Department of Physics, Harvard University, Cambridge MA 02138, USA}

\newcommand{\CfA}{Center for Astrophysics | Harvard \& Smithsonian, Cambridge, MA 02138, USA}

\newcommand{\NASASaganFellow}{NASA Sagan Fellow}

\newcommand{\UKRIFellow}{UKRI Future Leaders Fellow}

\newcommand{\UWIMadison}{Department of Astronomy, University of Wisconsin, Madison, WI 53706-1507, USA}

\newcommand{\DTUSpace}{DTU Space, National Space Institute, Technical University of Denmark, Elektrovej 328, DK-2800 Kgs. Lyngby, Denmark}

\newcommand{\GenevaObservatory}{Observatoire de Gen\`eve, Universit\'e de Gen\`eve, 51 chemin des Maillettes, 1290 Versoix, Switzerland}

\newcommand{\StAndrewsPhysicsAstronomy}{Centre for Exoplanet Science, SUPA, School of Physics and Astronomy, University of St Andrews, St Andrews KY16 9SS, UK}

\newcommand{\BelfastMathPhysics}{Astrophysics Research Centre, School of Mathematics and Physics, Queen's University Belfast, BT7 1NN, Belfast, UK}

\newcommand{\PadovaPhysicsAstronomy}{Dipartimento di Fisica e Astronomia ``Galileo Galilei'', Universit{\`a} di Padova, Vicolo dell`Osservatorio 3, I-35122 Padova, Italy}

\newcommand{\CavendishLab}{Astrophysics Group, Cavendish Laboratory, J.J. Thomson Avenue, Cambridge CB3 0HE, UK}

\newcommand{\INAFTorino}{INAF-Osservatorio Astrofisico di Torino, via Osservatorio 20, 10025 Pino Torinese, Italy}

\newcommand{\INAFPalermo}{INAF-Osservatorio Astronomico di Palermo, Piazza del Parlamento 1, 90134 Palermo, Italy}

\newcommand{\INAFBrenaBaja}{INAF-Fundacion Galileo Galilei, Rambla Jose Ana Fernandez Perez 7, E-38712 Brena Baja, Spain}

\newcommand{\MarylandPhysics}{Department of Physics, University of Maryland, College Park, MD 20742, USA}

\newcommand{\MarylandECE}{Department of Electrical and Computer Engineering, University of Maryland, College Park, MD 20742, USA}

\newcommand{\MarylandQTC}{Quantum Technology Center, University of Maryland, College Park, MD 20742, USA}

\newcommand{\KavliInstitute}{Kavli Institute for Cosmology, University of Cambridge, Madingley Road, Cambridge CB3 0HA, UK}

\newcommand{\Exeter}{Astrophysics Group, University of Exeter, Exeter EX4 2QL, UK}

\newcommand{\WarwickPhysics}{Department of Physics, University of Warwick, Coventry, CV4 7AL, UK}

\newcommand{\CEHWarwick}{Centre for Exoplanets and Habitability, University of Warwick, Coventry, CV4 7AL, UK}

\shorttitle{Estimating Magnetic Filling Factors From Simultaneous Spectroscopy and Photometry}
\shortauthors{T. Milbourne et al.}

\begin{document}

%%% TITLE %%%
\title{Estimating Magnetic Filling Factors From Simultaneous Spectroscopy and Photometry: Disentangling Spots, Plage, and Network}

%%% AUTHORS %%%
% TIER ONE
\author[0000-0001-5446-7712]{T. W. Milbourne}
\altaffiliation[Corresponding Author: ]{tmilbourne@g.harvard.edu}
\affiliation{\HarvardPhysics}
\affiliation{\CfA}

\author[0000-0001-5132-1339]{D. F. Phillips}
\affiliation{\CfA}

\author[0000-0003-2107-3308]{N. Langellier}

\affiliation{\HarvardPhysics}
\affiliation{\CfA}

\author[0000-0001-7254-4363]{A. Mortier}
\affiliation{\CavendishLab}
\affiliation{\KavliInstitute}

\author[0000-0001-9140-3574]{R. D. Haywood}
\altaffiliation{\NASASaganFellow}
\affiliation{\CfA}
\affiliation{\Exeter}

\author[0000-0001-7032-8480]{S. H. Saar}
\affiliation{\CfA}

%% TIER TWO

\author[0000-0001-8934-7315]{H. M. Cegla}
\altaffiliation{\UKRIFellow}
\affiliation{\CEHWarwick}
\affiliation{\WarwickPhysics}

\author[0000-0002-8863-7828]{A. Collier Cameron}
\affiliation{\StAndrewsPhysicsAstronomy}

%\author{J. Costes}
%\affiliation{\BelfastMathPhysics}

\author[0000-0002-9332-2011]{X. Dumusque}
\affiliation{\GenevaObservatory}

\author[0000-0001-9911-7388]{D. W. Latham}
\affiliation{\CfA}

\author[0000-0002-6492-2085]{L. Malavolta}
\affiliation{\PadovaPhysicsAstronomy}

\author[0000-0002-2218-5689]{J. Maldonado}
\affiliation{\INAFPalermo}

\author[0000-0002-8039-194X]{S. Thompson}
\affiliation{\CavendishLab}

\author{A. Vanderburg}
\affiliation{\UWIMadison}

\author{C. A. Watson}
\affiliation{\BelfastMathPhysics}

%TIER THREE

\author[0000-0003-1605-5666]{L. A. Buchhave}
\affiliation{\DTUSpace}

%\author[0000-0002-3697-1541]{N. Buchschacher}
%\affiliation{\GenevaObservatory}

\author[0000-0001-5701-2529]{M. Cecconi}
\affiliation{\INAFBrenaBaja}

%\author[0000-0002-9003-484X]{D. Charbonneau}
%\affiliation{\CfA}

\author[0000-0003-1784-1431]{R. Cosentino}
\affiliation{\INAFBrenaBaja}

\author[0000-0003-4702-5152]{A. Ghedina}
\affiliation{\INAFBrenaBaja}

%\author{A. G. Glenday}
%\affiliation{\CfA}

\author{M. Gonzalez}
\affiliation{\INAFBrenaBaja}

%\author{C-H. Li}
%\affiliation{\CfA}

\author[0000-0002-5432-9659]{M. Lodi}
\affiliation{\INAFBrenaBaja}

\author[0000-0003-3204-8183]{M. L\'opez-Morales}
\affiliation{\CfA}

%\author{C. Lovis}
%\affiliation{\GenevaObservatory}

%\author{M. Mayor}
%\affiliation{\GenevaObservatory}

%\author[0000-0002-9900-4751]{G. Micela}
%\affiliation{\INAFPalermo}

%\author[0000-0002-1742-7735]{E. Molinari}
%\affiliation{\INAFBrenaBaja}
%\affiliation{\INAFCagliari}

%\author{F. Pepe}
%\affiliation{\GenevaObservatory}

%\author{G. Piotto}
%\affiliation{\PadovaPhysicsAstronomy}
%\affiliation{\INAFPadova}

%\author[0000-0003-1200-0473]{E. Poretti}
%\affiliation{\INAFBrenaBaja}
%\affiliation{\INAFBrera}

%\author[0000-0002-6379-9185]{K. Rice}
%\affiliation{\EdinburghAstronomy}
%\affiliation{\EdinburghExoplanetCenter}

%\author[0000-0001-7014-1771]{D. Sasselov}
%\affiliation{\CfA}

%\author{D. S{\'e}gransan}
%\affiliation{\GenevaObservatory}

\author[0000-0002-7504-365X]{A. Sozzetti}
\affiliation{\INAFTorino}

%\author{A. Szentgyorgyi}
%\affiliation{\CfA}

%\author{S. Udry}
%\affiliation{\GenevaObservatory}

\author[0000-0003-0311-4751]{R. L. Walsworth}
\affiliation{\MarylandPhysics}
\affiliation{\MarylandECE}
\affiliation{\MarylandQTC}

%%% ABSTRACT %%%
\begin{abstract}
State of the art radial velocity (RV) exoplanet searches are limited by the effects of stellar magnetic activity. Magnetically active spots, plage, and network regions each have different impacts on the observed spectral lines, and therefore on the apparent stellar RV. Differentiating the relative coverage, or filling factors, of these active regions is thus necessary to differentiate between activity-driven RV signatures and Doppler shifts due to planetary orbits. In this work, we develop a technique to estimate feature-specific magnetic filling factors on stellar targets using only spectroscopic and photometric observations. We demonstrate linear and neural network implementations of our technique using observations from the solar telescope at HARPS-N, the HK Project at the Mt. Wilson Observatory, and the Total Irradiance Monitor onboard SORCE. We then compare the results of each technique to direct observations by the Solar Dynamics Observatory (SDO). Both implementations yield filling factor estimates that are highly correlated with the observed values. Modeling the solar RVs using these filling factors reproduces the expected contributions of the suppression of convective blueshift and rotational imbalance due to brightness inhomogeneities. Both implementations of this technique reduce the overall activity-driven RMS RVs from 1.64 m~s$^{-1}$ to 1.02 m~s$^{-1}$, corresponding to a 1.28 m~s$^{-1}$ reduction in the RMS variation. The technique provides an additional 0.41 m~s$^{-1}$ reduction in the RMS variation compared to traditional activity indicators.
\end{abstract}

%%% KEYWORDS %%%

%%% INTRODUCTION %%%
\section{Introduction}

\begin{figure*}
\begin{center} 
\includegraphics[width=.8\textwidth]{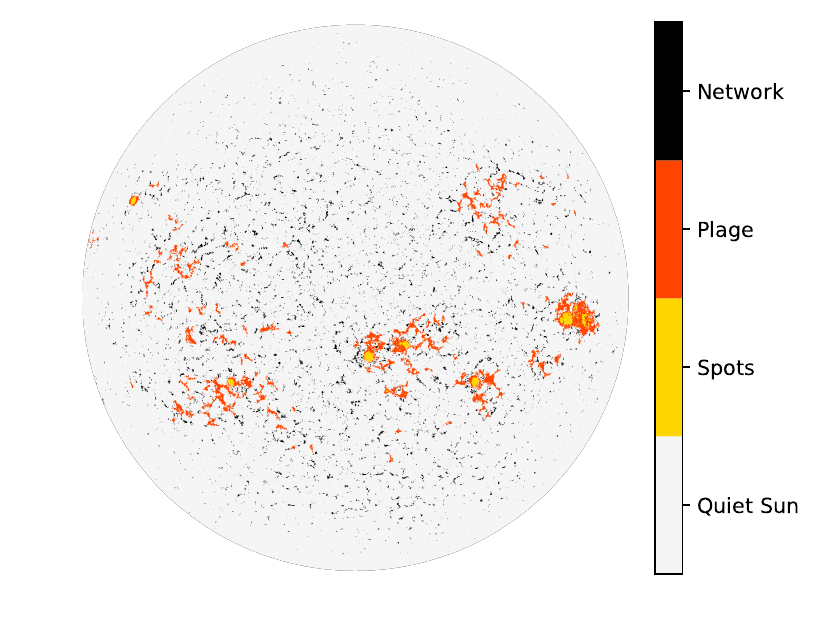} 
\caption{\label{fig:Thresholded}A representative HMI map of the three classes of active regions considered in this work. Spots, plage, and network are identified using the thresholding algorithm described by \cite{Haywood_2016} and MH19, with the threshold values given by \cite{Yeo_et_al_2013}. This algorithm is briefly recapped in Sec. \ref{sec:SDO}. Image taken January 1st, 2015 at 0:0:0.00 UT.}
\end{center} 
\end{figure*}

State of the art radial velocity (RV) searches for low-mass, long-period exoplanets are limited by the effects of stellar magnetic activity. An Earth-mass planet in the habitable zone of a Sun-like star has an RV amplitude on the order of 10 cm~s$^{-1}$. However, stellar activity processes on host stars, such as acoustic oscillations, magnetoconvection, suppression of convective blueshift, and long-term activity cycles, can produce signals with amplitudes exceeding 1 m~s$^{-1}$. A variety of techniques exist to mitigate the effect of these processes on the measured RVs: \cite{Chaplin_2019} discuss optimal exposure times to average out acoustic oscillations; \cite{Cegla2019} and \cite{Meunier_2017} present strategies for mitigaing the effects of granulation; and \cite{Aigrain2012}, \cite{Rajpaul2015}, \cite{haywood2020unsigned}, \cite{Langellier_2021}, and numerous others discuss statistically and physically-driven techniques for removing the effects of large-scale magnetic regions from RV measurements.

On timescales of the stellar rotation period, the apparent radial velocity is modulated by three main types of active regions: dark sunspots; large, bright plage; and small, bright network regions. These different regions may be identified using full-disk solar images, as shown in Fig.~\ref{fig:Thresholded}. \cite{Milbourne_2019} (hereafter referred to as MH19) found that the large-scale photospheric plage contribute differently to the solar suppression of convective blueshift than the smaller network. Failure to account for this different contribution leads to a significant RV shift over the 800 day span their of observations. Some of the long-term variation reported in MH19 may also be attributed to instrumental systematics: re-reducing the HARPS-N solar data with the ESPRESSO DRS \citep{Pepe2021, dumusque2020years} reduces this shift from  2.6 $\pm$ 0.3 m~s$^{-1}$ to 1.6 $\pm$ 0.5 m~s$^{-1}$. However, the remaining RV shift can only be fully removed by properly accounting for network regions in the calculated activity-driven RVs. While this analysis is possible on the Sun using high-resolution full disk images, traditional spectroscopic activity indicators, such as the Mt. Wilson S-Index \citep{Wilson_1968, Linksy_Avrett_1970} and the derivative index $\log(R'_{HK})$ \citep{Vaughan_1978,Noyes_et_al_1984} do not differentiate between large and small active regions. A new activity index or combination of activity indices is therefore necessary to successfully model the suppression of convective blueshift on stellar targets.

In this work, we demonstrate a new technique using simultaneous spectroscopy and photometry to estimate spot, plage, and network filling factors, and demonstrate that these filling factors may be used to model RV variations. In Section 2, we discuss the solar data used by our technique. An analytical implementation of the technique is described in Section 3, and a neural network implementation is presented in Section 4. The resulting solar filling factors, a model of the solar RVs, and possible applications to stellar targets are analyzed in Section 5.

\section{Measurements}

\begin{figure*} 
\begin{center} 
\includegraphics[width=.95\textwidth]{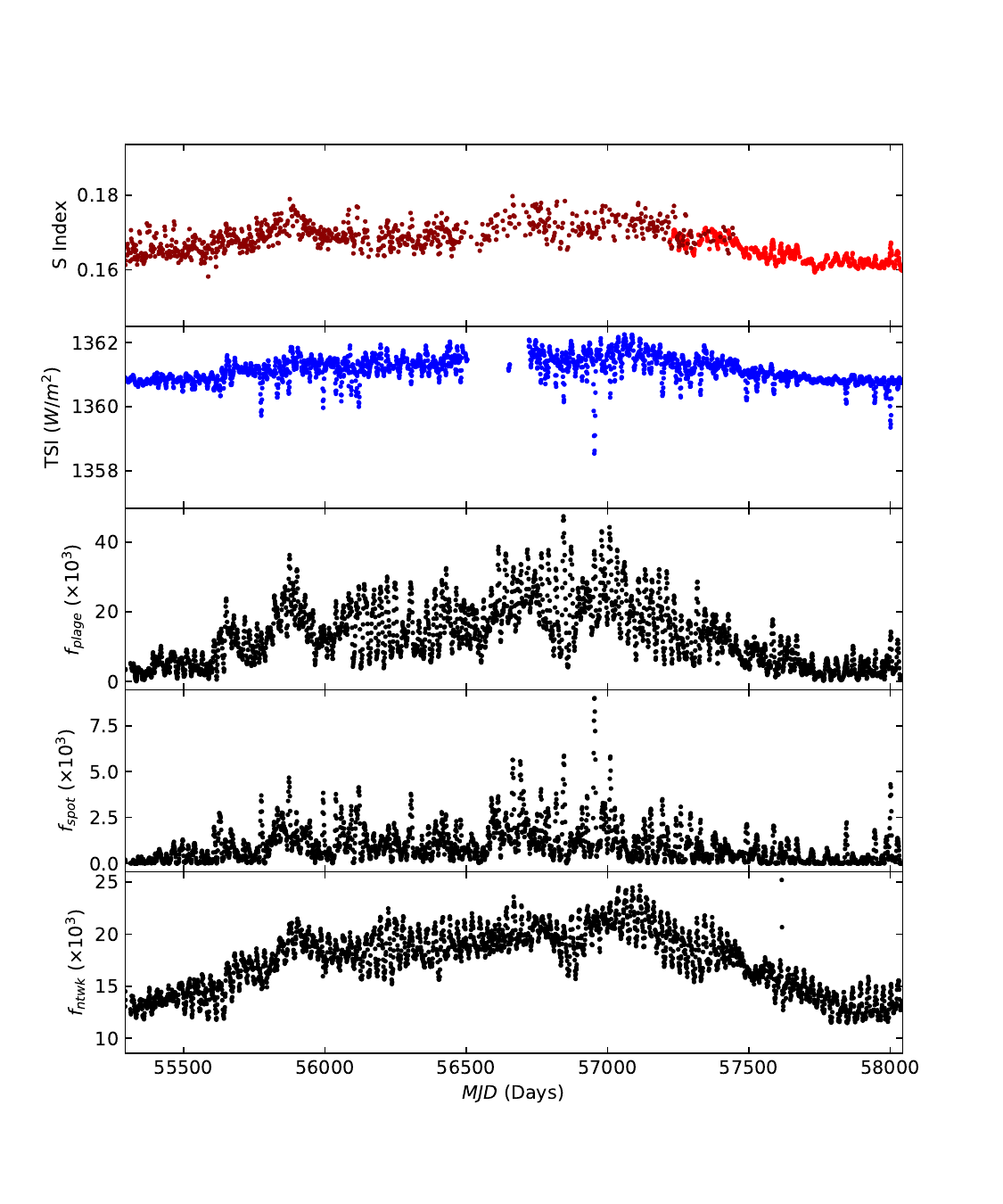} 
\caption{\label{fig:AllVars}Time series of solar observations used in this work. From top to bottom: Mt.\ Wilson and HARPS-N solar telescope observations of the calcium S-index (red); Total Solar Irradiance (TSI) from SORCE/TIM (blue); and SDO/HMI plage, spot, and network filling factors (black). Note that the consistent overall shapes of the S-index, TSI, and bright filling factors, and that dips in the TSI are coincident with peaks in the spot filling factor. Observations are taken between April 2010 through October 2017. Note the two different reds used in the S-index plot: the darker red points correspond to measurements by Mt. Wilson, and bright red points are from the HARPS-N solar telescope. Note that due to an instrumental anomaly, no TSI data is availble from SORCE/TIM from mid July 2013 until March 2014 - we therefore do not use any times in this period in our analysis \citep{Kopp2005a, Kopp2005b}}.

\end{center} 
\end{figure*}

\subsection{HARPS-N/Mt. Wilson Survey}

We use the HARPS-N solar telescope \citep{Dumusque:2015, Phillips2016} measurements of the S-index, as described in MH19 and \cite{ACC2019}. The S-index quantitatively represents activity-driven chromospheric re-emission in the Calcium II H and K lines. The presence of spots, plage, and network all increase the S-index. The solar telescope takes exposures every five minutes while the Sun is visible. Each measurement of the S-index has an average precision of $2.5 \times 10^{-4}$, or a fractional uncertainty of 0.0016.

Note that  HARPS-N solar telescope observations began in mid 2015. To cover the rest of the solar cycle, we use data from the Mt.\ Wilson S-index survey, as presented by \cite{Egeland_2017}. Observations from the two instruments overlap between July 2015 and February 2016 (JD 2457222 and JD 2457444), allowing us to combine these time series. The solar telescope dataset is rescaled so that the points in the overlapping time interval have the same mean and variance as the Mt. Wilson data from the same time interval, as described in \cite{haywood2020unsigned}. The resulting combined dataset is shown in the top panel of Fig.~\ref{fig:AllVars}.

\subsubsection{HARPS-N Solar RVs}
We use the HARPS-N solar telescope's measurements of the solar RVs to assess our ability to model realistic RV variations using our estimated filling factors. \footnote{We use publicly available HARPS-N solar telescope observations reduced using the most recent ESPRESSO pipeline, available at \url{https://dace.unige.ch/sun/}} \citep{Phillips2016, ACC2019, dumusque2020years}, as well as our estimated values derived from the linear and MLP techniques. The HARPS-N RVs used in this work span the period from July 2015 to October 2017, with exposures taken every five minutes while the Sun is visible. Each RV measurement has an average precision of 23 cm~s$^{-1}$.
\subsection{SORCE}

We use the Total Irradiance Monitor (TIM) onboard the Solar Radiation and Climate Experiment (SORCE) \citep{Rottman2005,Kopp2005a, Kopp2005b} to measure photometry for the whole solar cycle. The total solar irradiance (TSI) is the solar analogue of the light curves obtained by \emph{Kepler}, K2, TESS, and CHEOPS \citep{Borucki977, 1538-3873-126-938-398, TESS2014, CHEOPS2017}, though the Sun's proximity means it can be observed continuously over much longer periods. The TIM level 3 data products are averaged over 6 hours, with a precision of 0.005 $W/m^2$.

We expect the overall brightness of the Sun to vary with the stellar cycle. Its relative brightness increases with the presence of plage and network, and decreases with the presence of spots. This modulation makes the TSI, and stellar light curves in general, useful tools for isolating the effects of stellar magnetic activity \citep{Aigrain2012}. The time series of the TSI is shown in the second panel of Fig. \ref{fig:AllVars}.

\subsection{SDO}
\label{sec:SDO}
We use images from the Helioseismic and Magnetic Imager (HMI) instrument onboard the Solar Dynamics Observatory (SDO, \citealt{Schou2012,Pesnell2012,Couvidat2016}) to independently calculate solar filling factors. HMI measures the 6173.3 \AA\ iron line at six points in wavelength space using two polarizations. From these measurements, they reconstruct the Doppler shift and magnetic field strength along with the continuum intensity, line width, and line depth at each point on the solar disk.

Spots, plage, and network are identified on HMI images using a simple threshold algorithm:

\begin{itemize}
    \item An HMI pixel is considered magnetically active if the radial component of the magnetic field is over three times greater than the expected noise floor: $|B_{r}| > 3\sigma_{B_r}$. 
    \item Active pixels below the intensity threshold of \cite{Yeo_et_al_2013}, such that $I < 0.89I_{quiet}$, are labelled as spots. Here, $I_{quiet}$ is the average intensity of inactive pixels on a given image.
    \item Active regions exceeding the above intensity threshold that span an area > 20 micro-hemispheres (that is, 20 parts per million of the visible hemisphere), or $60 \ \rm Mm^2$, are labelled as plage.
    \item Active regions exceeding the intensity threshold that span an area < 20 micro-hemispheres are labelled as network.
\end{itemize}
These calculations are explained in further detail in MH19. %Fig.~\ref{fig:Thresholded} shows a representative thresholded HMI image, and 
The resulting filling factors for each feature are plotted in the bottom three panels of Fig.~\ref{fig:AllVars}. We use one HMI image taken every four hours in our analysis. The photon noise at disk center for the magnetograms and continuum intensity for these HMI images are $\sigma_{B_r} = 8$ G and $\sigma_{I_c} = 0.01\%$ respectively \citep{Couvidat2016}. This corresponds to uncertainties $< 0.1\%$ in the resulting magnetic filling factors.

Since SDO/HMI allows us to perform precise direct, independent measurements of the three filling factors of interest, we use these results as the "ground truth" in our analysis.

An SDO analogue does not exist for non-solar stars, so if we wish to determine feature-specific filling factors of stellar targets we must make indirect estimates of the filling factors using spectroscopic and photometric data. In the next section, we discuss two processes to do so. To mitigate the effects of acoustic oscillations, granulation, and other short-timescale activity process, we take daily averages of each set of observations used in our analysis. We also interpolate the HARPS-N/Mt. Wilson observations, SORCE/TIM observations, and SDO/HMI filling factors onto a common time grid of one observation each day, when all three instruments have measurements.

\section{Methods}

In this section, we present analytical and neural net implementations of a technique to determine the spot, plage, and magnetic filling factors using only spectroscopic and photometric data. The analytical implementation (hereafter referred to as the"linear technique") only requires knowledge of the star's distance and radius, along with estimates of the spot and plage/network temperature contrasts and the quiet star effective temperature. The neural net implementation infers these parameters from the data, and therefore only requires spectroscopic and photometric observations of the target. This technique can therefore be used to estimate filling factors without prior knowledge of the filling factors from full-disk images. Furthermore, if the resulting filling factors are only being used to model activity-driven RV variations, only time series correlated with the spot, network, and plage filling factors are needed, further simplifying the required knowledge of the star.

\subsection{Filling Factor Modelling: Linear Technique}

\subsubsection{Modelling Irradiance Variations Using Filling Factors}

In MH19, the authors reproduce the observed TSI using a linear combination of the spot and plage filling factors. Following \cite{Meunier_et_al_2010}, they use the SDO/HMI derived plage and spot filling factors, and assume that the solar irradiance follows the Stefan-Boltzmann law for blackbodies:

\begin{multline}
\label{eq:TSImodel}
    TSI = \mathcal{A} \sigma \left[(1-f_{spot}-f_{bright})T_{quiet}^4 \right.
    \\+ f_{spot}(T_{quiet} + \Delta T_{spot})^4
    \\+ \left. f_{bright}(T_{quiet}+\Delta T_{bright})^4 \right]
\end{multline}

\noindent where $\sigma$ is the Stefan-Boltzmann constant, $\mathcal{A}= (R_{\odot}/1\ \mathrm{AU})^2$ is a geometrical constant relating the energy emitted at the solar surface to the energy received at Earth, $T_{quiet}$ is the quiet Sun temperature, $\Delta T_{spot}$ and $\Delta T_{bright}$ are the effective temperature contrasts of spots and plage/network regions, and $f_{spot}$ and $f_{bright}$ are the HMI spot and plage/network filling factors. Expanding as a power series yields the following approximation:

\begin{equation}
\label{eq:TSImodelsimp}
    TSI \approx \mathcal{A} \sigma T_{quiet}^4 \left(1 + 4 \frac{\Delta T_{spot}}{T_{quiet}}f_{spot} + 4 \frac{\Delta T_{bright}}{T_{quiet}} f_{bright} \right)
\end{equation}

In MH19, the authors show that HMI observations of filling factors may be used to reproduce SORCE TSI given temperature contrasts for plage/network features and spots and the effective temperature of the quiet Sun. In this work, we invert the process and use the resulting effective temperatures of each type of active region, along with the correlations between the TSI, S index, and filling factors demonstrated in MH19, to reproduce the observed magnetic filling factors for each type of active region. The potential stellar applications of this technique are discussed in more detail in Sec. \ref{sec:StellarOutlook}.

We begin by fitting the SORCE TSI to Eq. \ref{eq:TSImodelsimp} using the SDO/HMI measurements of $f_{spot}$ and $f_{bright}$. This fit yields $T_{quiet} = 5769.85 \pm 0.01$ K, $\Delta T_{spot} = -525 \pm 8$ K, and $\Delta T_{bright} = 46.1 \pm 0.5$ K. Also note that the solar radius varies as a function of wavelength. To be consistent with HMI, we use $R_{\odot} = 695982 \pm 13$ km, the solar radius measured at 6173.3 \AA\ \citep{Rozelot_2015}.

\begin{figure}
\begin{center} 
\includegraphics[width=\columnwidth]{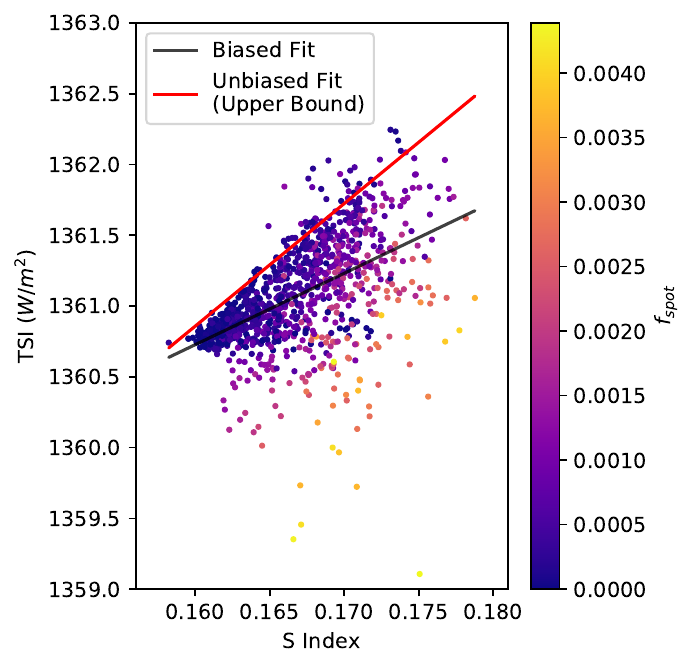} 
\caption{\label{fig:SlopeFit}A plot of the SORCE TSI versus the HARPS-N/Mt. Wilson S index. The color of each point corresponds to the value of $f_{spot}$. We see that the S index is highly correlated with the TSI, as expected. We may use this correlation to estimate the plage/network filling factors on the Sun. However, increased spot coverage results in a lower TSI value for a given S index, which will bias our estimate of bright region filling factor. This, in turn, will result in a less accurate estimate of the spot filling factor. The black line shows the result of the straightforward linear fit of TSI and S index, which is biased as described above. To isolate the plage and network driven TSI variations, we find the 50\% most densely clustered points in the above scatter plot, and fit a line to the upper boundary of this region. (This choice in point density is arbitrary, but the resulting best-fit line is robust to variations in this parameter.) The resulting fit line, shown in red, is unbiased by the presence of spots.}
\end{center} 
\end{figure}

\subsubsection{Differentiating Bright and Dark Regions}
 The brightness of a Sun-like star may be modulated by the long-term stellar activity cycle (\emph{e.g.}, the 11-year solar cycle). Since $\Delta T_{bright} > 0$ and $\Delta T_{spot} < 0$, we see from Eq.~\ref{eq:TSImodelsimp} that TSI variations (on timescales of the rotation period) below the value expected from the activity cycle must be the result of spots. Since the Sun is plage dominated (that is, $f_{bright} > 100 f_{spot}$, as shown in \citealt{Shapiro_et_al_2015}, MH19, and Fig. \ref{fig:AllVars}), plage and network are the primary source of variation of the TSI and S-index, with spots making negligible contributions to the variability of the irradiance on timescales of the solar cycle. This is also visible in comparing the plots of the TSI, S index, $f_{plage}$, and $f_{ntwk}$ shown in Fig. \ref{fig:AllVars}, and is also discussed in detail by MH19. We may therefore use a linear transformation of the S-index to provide an initial estimate of $f_{bright}$, and then use the TSI to estimate $f_{spot}$. The full calculation is as follows:

(1) We begin by assuming that the S-index is directly proportional to the total plage and network filling factor (that is, $f_{bright, 1} = m_1 S_{HK} + b_1$), and that the plage and network are the dominant drivers of TSI variation. Our first estimate of the spot filling factor is therefore $f_{spot, 1}= 0$. We then estimate the values of $m_1$ and $b_1$ by fitting the TSI as a linear transformation of the S-index:
\begin{equation}
\label{eq:TSIfitstep1}
TSI_{fit,1}(m_1, b_1) = \mathcal{A} \sigma T_{quiet}^4 \left(1 + 4 \frac{\Delta T_{bright}}{T_{quiet}} \left( m_1 S_{HK} + b_1 \right) \right).
\end{equation}
Note that we have included the physical constants for normalization though they are degenerate with the fit parameters.

It is not sufficient to perform a simple linear fit to the TSI and S index. In the above step, we model the activity-driven variations of the TSI due to the presence of bright regions. However, as shown in Fig.~\ref{fig:SlopeFit}, the presence of spots produces scatter in this relationship. This scatter is in one direction: as $f_{spot}$ increases, the observed TSI for a given value of the S-index decreases. To isolate the activity-driven TSI variations due to $f_{bright}$, we determine the 50\% most densely clustered points in Fig.~\ref{fig:SlopeFit}, and fit the upper boundary of this region. The best fit line of this upper bound gives us the values of $m_1$ and $b_1$ used above.

(2) Next, we assume any deviation from the fit above are driven by spots, which are not included in this model. We then make a second estimate of the spot filling factor, $f_{spot,2}$, from the residuals to the above fit:
\begin{equation}
\label{eq:TSIfitstep1p5}
    f_{spot, 2} =
    \begin{dcases} 
          \frac{TSI - TSI_{fit,1}(m_1, b_1)}{4 \mathcal{A} \sigma \Delta T_{spot}  T_{quiet}^3}, & TSI - TSI_{fit,1}\leq 0 \\
          0, & TSI - TSI_{fit,1} > 0.
   \end{dcases}
\end{equation}
Essentially, any point below the line of best fit in Fig.~\ref{fig:SlopeFit} is assumed to be due to spot-driven brightness variations. This increases the importance of avoiding spot-driven biases in Step 1. If a simple linear fit is used in Step 1 instead of the fit to the upper boundary described above, the presence of spots will reduce the slope of the best-fit line, which will result in an artificially reduced $f_{bright}$ value, and will also exclude real spot-driven variations from our calculation of $f_{spot, 2}$.

(3) We determine our final estimate of $f_{bright}$ and $f_{spot}$ by fitting the following expression to the TSI:
\begin{multline}
\label{eq:TSIfitstep2}
TSI_{fit,2} = \mathcal{A} \sigma T_{quiet}^4 \left(1 +4 \frac{\Delta T_{spot}}{T_{quiet}}\left( a_2 f_{spot, 2} \right) \right. \\ + \left. 4 \frac{\Delta T_{bright}}{T_{quiet}} \left( m_2 S_{HK} + b_2 \right) \right).
\end{multline}
\noindent where our estimated values of $f_{bright}$ and $f_{spot}$ are given by

\begin{equation}
    \label{eq:LinearBright}
    f_{bright} = m_2 S_{HK} + b_2
\end{equation}

\begin{equation}
    \label{eq:LinearSpot}
f_{spot} = a_2 f_{spot,2}
\end{equation}

\noindent and the parameters $a_2$, $m_2$, and $b_2$ are determined by the above fit. The resulting best-fit parameters derived from the solar case are given in Table \ref{tab:LinearParam}. Note that we do not expect $m_2$ and $b_2$ to be very different from the parameters $m_1$ and $b_1$ found previously, nor do we expect $a_2$ to be very different from 1. However, since we exclude any negative residuals from our estimate of $f_{spot}$ in Step 2 above, we perform this final fit in case this excluded information changes the best-fit parameters in any way.

As previously noted, this technique only requires knowledge of the star's distance and radius, along with estimates of the spot and plage/network temperature contrasts and the quiet star effective temperature - this means that it can be used to estimate filling factors without prior knowledge of the filling factors from full-disk images. If the plage and network features are only being used to decorrelate activity-driven RV variations, only time series correlated with the spot, network, and plage filling factors are needed, and the above terms may be absorbed into the fit coefficients in Eqs.~\ref{eq:TSIfitstep1} and \ref{eq:TSIfitstep2}---this is discussed further in Sec.~\ref{sec:StellarOutlook}.

\begin{center}
\begin{table}
\centering
\begin{tabular}{c|c}
    $m_1$ & $2.02 \pm 0.07$\\
    $b_1$ & $-0.31 \pm 0.01$ \\
    $a_2$ & $0.9912 \pm 0.0008$\\
    $m_2$ & $2.072 \pm 0.002$\\
    $b_2$ & $-0.3169 \pm 0.0003$ \\
\end{tabular}
\caption{Best-fit parameters for the linear filling factor estimation technique. As expected, $m_1$ and $m_2$ are consistent within error bars, as are $b_1$ and $b_2$. Similarly, $a_2$ is very close to 1, as expected..}
\label{tab:LinearParam}
\end{table}
\end{center}

\subsubsection{Differentiating the Network and Plage Filling Factor}

In the discussion above, we extract $f_{bright}$, the combined plage and network filling factor. However, we may consider these two separately by adding a network term to Eq.~\ref{eq:TSImodel}:
\begin{multline}
\label{eq:TSImodelnet}
TSI = \mathcal{A} \sigma \left[(1-f_{spot}-f_{plage}-f_{ntwk})T_{quiet}^4 \right. \\
+ f_{spot}(T_{quiet} + \Delta T_{spot})^4  \\
+ f_{plage}(T_{quiet}+\Delta T_{plage})^4 \\
+ \left. f_{ntwk}(T_{quiet}+\Delta T_{ntwk})^4\right].
\end{multline}
Fitting this equation to the TSI, this time using the SDO observed filling factors  as inputs, reveals that the plage and network have distinct effective temperatures, $\Delta T_{plage} = 32 \pm 1 K$ and $\Delta T_{ntwk} = 79 \pm 2 K$. This is consistent with the intensity maps produced by HMI, which show that network regions are, on average, indeed brighter than plage These temperature contrasts are necessary for separating the plage and network contributions to the filling factor.

Setting Eq.~\ref{eq:TSImodel} equal to Eq.~\ref{eq:TSImodelnet} and expanding as a power series, we find $$f_{bright}\Delta T_{bright} \approx f_{plage}\Delta T_{plage} + f_{ntwk}\Delta T_{ntwk}.$$ Since areas are additive, we also expect $$f_{bright} = f_{plage} + f_{ntwk}.$$
Combining these equations and solving for $f_{ntwk}$ and $f_{plage}$ in terms of $f_{bright}$ yields the following expressions:
\begin{equation}
\label{eq:LinearNtwk}
 f_{ntwk} = \frac{\Delta T_{plage} - \Delta T_{bright}}{\Delta T_{plage} - \Delta T_{ntwk}} f_{bright} + \mathcal{B}, \\
\end{equation}
and
\begin{equation}
\label{eq:LinearPlage}
f_{plage} = \frac{\Delta T_{bright}- \Delta T_{ntwk}}{\Delta T_{plage}- \Delta T_{ntwk}} f_{bright} - \mathcal{B}.
\end{equation}
Note the prefactors for each estimate, which simply rescale the brightness contributions of each class of active region to account for the different effective temperatures. Also note the offset $\mathcal{B}$ in our estimate of $f_{plage}$: this accounts for the fact that $f_{plage}$ goes to 0 at solar minimum, while $f_{ntwk}$ has a basal value at solar minimum. In this analysis, the value of $\mathcal{B}$ may be found from the expected value of $f_{bright}$ at solar minimum:
\begin{equation}
    \mathcal{B} = \frac{\Delta T_{bright}- \Delta T_{ntwk}}{\Delta T_{plage}- \Delta T_{ntwk}} \textrm{min} \left(f_{bright}\right).
\end{equation}
Determining this offset therefore requires TSI and S index observations taken at solar minimum, which may increase the observational load associated with this technique. However, modeling the effects of network on the activity-driven solar RV variations only requires a quantity \textit{correlated with} $f_{plage}$. The value of this offset is therefore unimportant for our purposes.

\subsection{Filling Factor Modelling: Machine Learning Technique}

While machine learning techniques are predicatively powerful, their black-box nature makes them not physically explanatory, and therefore not necessarily useful for some scientific applications. However, the existence of a clear causal connection between the S-index, TSI, and filling factors makes machine learning a strong candidate for the problem of estimating feature-specific magnetic filling factors from spectroscopic and photometric information. We already know the physics connecting these variables, and can therefore have machine learning "discover" and refine the relationships found above. 
%The machine learning approach 
A neural network used as a universal function approximator~\citep{Cybenko:1989}
may be able to determine subtle details of these relationships that are not incorporated into our linear model, such as the different effects of network vs. plage, how underlying spatial distributions of active regions affect the resulting filling factors and activity indicators, and the correlations between spots and plage.
\begin{center}
\begin{table}
\centering
\begin{tabular}{c|c|c}
    Hidden Layer Sizes&$\alpha$ & $\beta$ \\
     \hline
    (64,64) & 0.0001 & 0.001 \\
\end{tabular}
\caption{Hyperparameter values for MLP filling factor calculation as optimized from cross-validation. Here $\alpha$ gives the $L_2$ regularization parameter and $\beta$ is the learning rate.}
\label{tab:NNparam1}
\end{table}
\end{center}
We therefore compare the linear technique discussed in the previous section with a type of neural network known as a Multilayer Perceptron (MLP, \citealt{HINTON1989185}). The MLP consists of an input layer, several fully-connected hidden layers, and an output layer. It is one of the simplest neural networks that may be used as a universal function approximator, making it ideal for this application. We implement the MLP using the \texttt{MLPRegressor} class in the \texttt{scikit-learn} package in Python \citep{Python, scikit-learn}.

We train the MLP using the TSI and S-index inputs, and using the SDO plage, spot, and network filling factors as outputs. 75\% of the total available data (taken over the whole solar cycle) is used for training, with 25\% set aside to test the performance of the trained network. The MLP uses two hidden network layers, each with 64 neurons. We optimized the size and number of these layers as well as the $L_2$ regularization parameter, $\alpha$, which combats overfitting by constraining the size of the fit parameters as measured with an $L_2$ norm; and the learning rate, $\beta$, which controls the step-size in the parameter space search using five-fold cross validation. That is, we randomly shuffled the training data and divided it into five groups. We trained the network on four of these groups, and then tested the network on the remaining group. We repeated this process using each of the five groups as a test set to mitigate the effects of overfitting on our network, and then repeated the entire five-fold process using each combination of network parameters to determine which combination of hyperparameters resulted in the best performance. The resulting values are summarized in Table \ref{tab:NNparam1}. The network was optimized to minimize square error using a stochastic gradient descent algorithm (SGD), and was trained for a maximum of $10^4$ steps (though the algorithm may stop training earlier once the network converges.)

Note that we may also use this MLP approach to fit the solar RVs directly using the TSI and S-index, without first computing magnetic filling factors. In Sec. \ref{sec:RVapp}, we compare a direct MLP fit of this form to RV models dervived from our estimated filling factors to determine if there is any additional RV information in the TSI and S index which is not incorporated into our filling factor estimates.

\subsection{Radial Velocity Modelling}

\begin{table}
\centering
\begin{tabular}{c|c}
Filling Factor Source & $r\left(\Delta v_{conv}, RV(f_{spot}, f_{ntwk}, f_{plage})\right)$ \\
 \hline
HMI  & 0.92 \\
Linear Estimate & 0.84 \\
MLP Estimate & 0.83
\end{tabular}
\caption{Pearson correlation coefficients between HMI derived estimate of the suppression of convective blueshift, $\Delta v_{conv}$ and the activity driven RVs derived from Eq.~\ref{eq:vcon}. The very high correlation coefficients indicate that the plage and network filling factors successfully estimate the RV contribution of the suppression of convective blueshift, as expected from MH19.}
\label{tab:vcon_corr}
\end{table}

\begin{table}
\centering
\begin{tabular}{c|c}
Filling Factor Source & $r\left(\Delta v_{phot}, RV(f_{spot})\right)$ \\
 \hline
HMI  & 0.62 \\
Linear Estimate & 0.70 \\
MLP Estimate & 0.67
\end{tabular}
\caption{Pearson correlation coefficients between HMI derived estimate of the photometric velocity shift, $\Delta v_{phot}$ and the activity driven RVs derived from Eq.~\ref{eq:vphot}. The relatively high correlation coefficients indicate that the spot filling factors successfully estimate the photometric RV shifts, as expected from MH19.}
\label{tab:vphot_corr}
\end{table}

\begin{table}
\begin{tabular}{c|c}
 & RMS (m~s$^{-1}$) \\
 \hline
 Full solar dataset & 1.64\\
 Decorrelated with S index & 1.10 \\
 Decorrelated with HMI filling factors & 0.91\\
 Decorrelated with linear filling factor estimates & 1.04\\
 Decorrelated with MLP filling factor estimates  & 1.02\\
 Decorrelated with MLP RV estimate & 0.96\\

\end{tabular}
\caption{RMS RV residuals from several models and methods. Using our estimates of $f_{spot}$, $f_{plage}$, and $f_{ntwk}$ in Eq. \ref{eq:RVmodel} reduces the RMS RVs by 60 cm~s$^{-1}$. However, using the HMI-observed filling factors reduces the RMS residuals by a further 13 cm~s$^{-1}$, indicating there is additional information in these filling factors not captured by our estimates. A direct MLP fit to the solar RVs, using the S index and TSI as inputs, performs better than our estimated filling factors, but does not perform as well as the fit to HMI filling factors. This indicates that, while our estimated filling factors are highly correlated with the observed values, the S index and TSI alone are insufficient to completely characterize the filling factors of each feature.}
\label{tab:RVrms}
\end{table}

In MH19, the authors found that the HARPS-N solar radial velocities were well-represented by a linear combination of $\Delta v_{conv}$, the suppression of convective blueshift, and $\Delta v_{phot}$, the photometric velocity shift due to bright and dark active regions breaking the symmetry of the solar rotational profile:
\begin{equation}
    \label{eq:RVmodel_MH2019}
    RV = A_1\Delta v_{phot} + B_1 \Delta v_{conv} + RV_0.
\end{equation}
Here, we see if we can perform a similar reconstruction using our estimates of the magnetic filling factors. Since the presence of active regions drives the suppression of convective blueshift, we expect the $\Delta v_{conv}$ to be proportional to the spot, plage, and network filling factors. Based on the results of MH19, we also expect network and plage regions to have different contributions to $\Delta v_{conv}$. We therefore model the suppression of convective blueshift as:

\begin{equation}
\label{eq:vcon}
    \Delta v_{conv} = B f_{spot} + C f_{plage} + D f_{network} + E.
\end{equation}

While plage and network occupy a greater area than spots on the Sun, and therefore dominate the suppression of convective blueshift, the higher brightness contrast of spots means that they drive the photometric RV shift, $\Delta v_{phot}$. We expect  $\Delta v_{phot}$ to scale with number and size of the spots rotating across the solar surface. However, we also expect a phase lag between $f_{spot}$ and $\Delta v_{phot}$. For a single spot moving across the solar disk, $f_{spot}$ is at its maximum value when the spot is on the center of the solar disk. However, the absolute value of $\Delta v_{phot}$ is maximized when the spot is at the solar limb, rotating toward or away from the observer, and is zero when the spot is at disk center. We therefore expect $\Delta v_{phot}$ to also depend on the derivative of the filling factor with respect to time:
\begin{equation}
\label{eq:vphot}
    \Delta v_{phot} \propto f_{spot} \times \left(\frac{df_{spot}}{dt}\right).
\end{equation}
Note that this formulation mirrors the FF$^\prime$ method developed by \cite{Aigrain2012}.

By combining Eqs. \ref{eq:RVmodel_MH2019}, \ref{eq:vcon}, and \ref{eq:vphot}, we therefore produce a model of the solar RVs based on our estimated feature-specific magnetic filling factors:
\begin{equation}
    \label{eq:RVmodel}
    RV = A f_{spot} \left(\frac{df_{spot}}{dt}\right) + B f_{spot} + C f_{plage} + D f_{network} + RV_0.
\end{equation} 

\noindent Note that the offset $E$ in Eq. \ref{eq:vcon} has been absorbed into $RV_0$.

Fitting to the HMI-observed filling factors reduces the HARPS-N solar RV residuals from 1.64 m~s$^{-1}$ to 0.91 m~s$^{-1}$, as shown in Table \ref{tab:RVrms}. In comparison, the usual technique of simply decorrelating the S-index from the RV measurements (i.e., fitting $RV = A S_{HK} + B$) results in an RMS of only 1.10 m~s$^{-1}$, indicating that spots, plage, and network regions have different contributions to the S index, and have different effects on the suppression of convective blueshift \citep[][Fig.7]{Meunier2010b}. 

Repeating this fit with both our linear and MLP estimates of $f_{spot}$, $f_{ntwk}$, and $f_{plage}$ reduces the RMS RV to 1.04 m~s$^{-1}$ and 1.02 m~s$^{-1}$ respectively. This implies that, while our estimates are highly correlated with the true values of the filling factors, there is additional information in the true filling factors that is not captured by either technique, resulting in less precise estimates of the convective blueshift and photometric RV shifts. Interestingly, while the linear filling factor estimates cannot distinguish the RV contributions of the spots and network, as discussed above, the linear and MLP estimates result in similar RMS RVs. The fact that the linear estimates reduce the RMS RVs below the level obtained from the S index despite this limitation highlights the importance of spots in our models of activity-driven RVs.

%%% DISCUSSION %%%
\section{Results}

\subsection{Spot, Plage, and Network Filling Factors}

\begin{figure*} 
\begin{center} 
\includegraphics[width=1\textwidth]{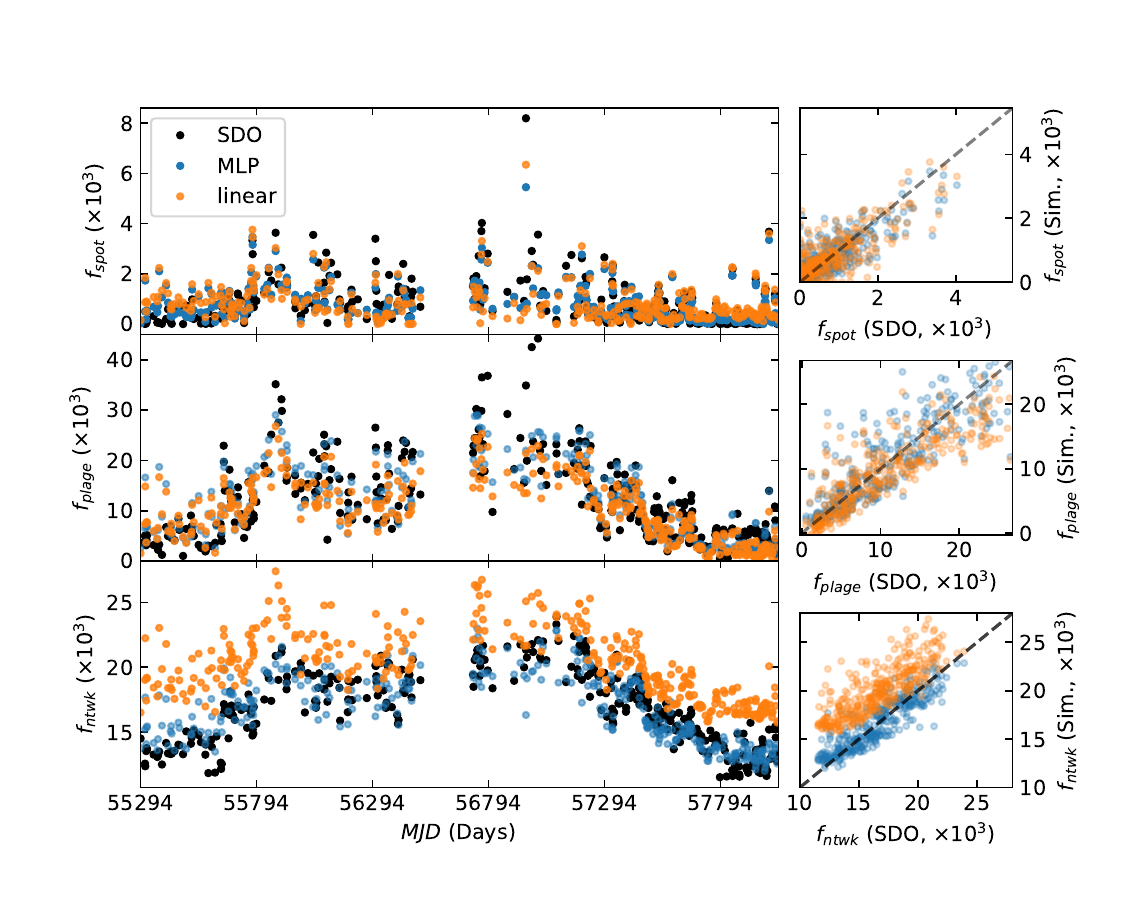} 
\caption{\label{fig:Results}Comparison of the SDO/HMI-measured magnetic filling factors (black) to the machine learning (blue) and linear (orange) estimates derived from the S-index and TSI. The time series for the three filling factors are plotted in the left column. The estimated filling factors are plotted as a function of the HMI filling factors in the right column---the grey dashed lines indicate a slope of 1, and are meant to guide the eye. Both the linear and machine learning techniques reproduce the directly-observed values of $f_{spot}$, $f_{plage}$, and $f_{ntwk}$. Note that there is a slight offset between the linear estimate of $f_{ntwk}$ and the SDO measurements. However, this offset is well within the expected 20\% - 50\% definitional variations reported by \cite{Meunier_et_al_2010}
}
\end{center} 
\end{figure*}

\begin{figure*}
\begin{center} 
\centering
\includegraphics[width=1\textwidth]{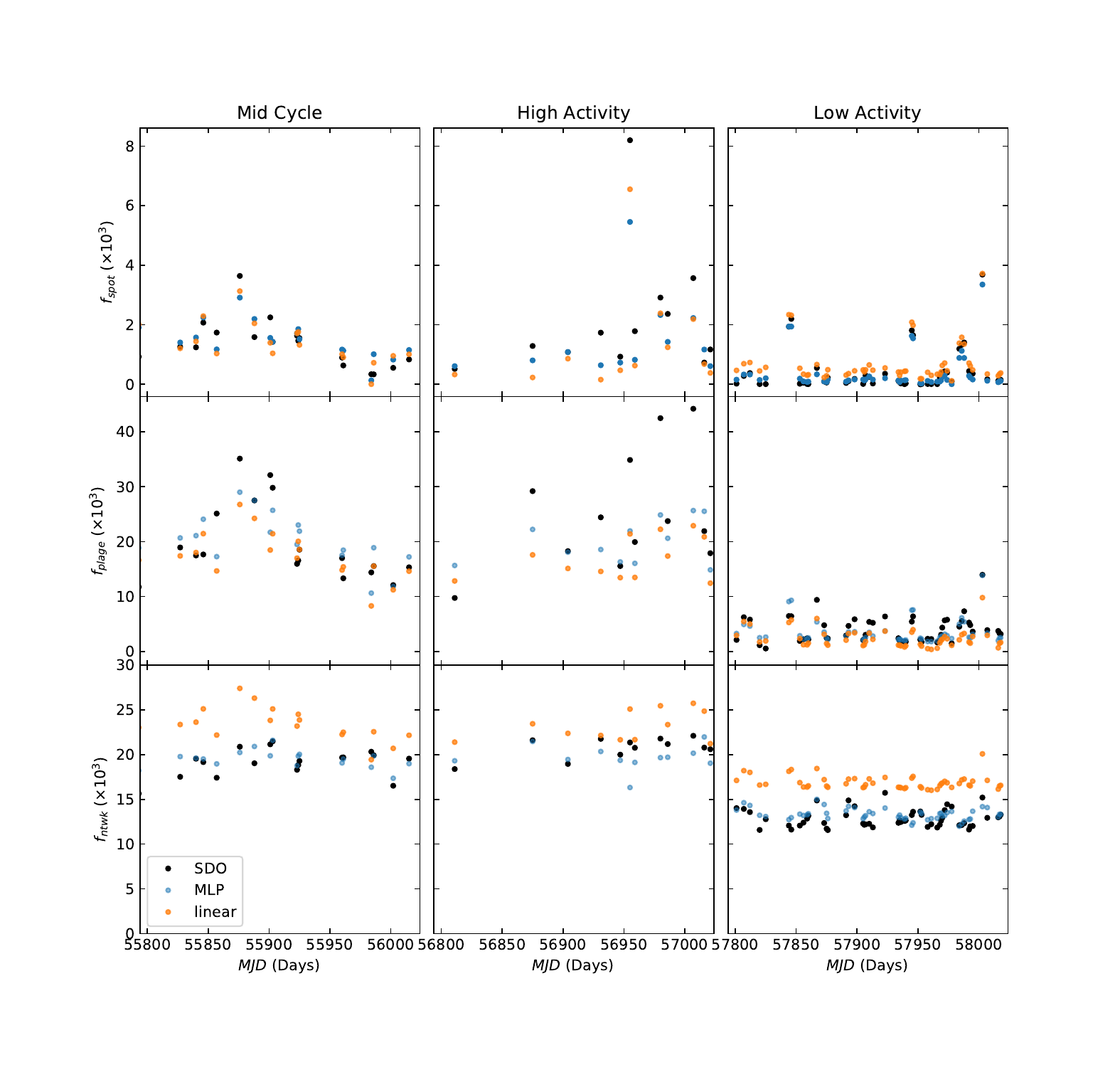}
\caption{\label{fig:ResultsZoom}230 day subsets of the time series of the SDO/HMI-observed magnetic filling factors (black), along with the MLP (blue) and linear (orange) estimates derived from the S-index and TSI. Three subsets are shown, taken during the middle of the stellar cycle (left), during solar maximum (middle), and approaching solar minimum (right). Both techniques successfully reproduce $f_{spot}$, $f_{plage}$, and $f_{ntwk}$, with especially good performance at solar minimum.
}
\end{center}
\end{figure*}

Fig.~\ref{fig:Results} shows that both the linear and MLP-based techniques successfully reproduce the directly-observed values of $f_{spot}$, $f_{plage}$, and $f_{ntwk}$.  Fig.~\ref{fig:ResultsZoom} shows the same information as Fig.~\ref{fig:Results}, but for three 230 day regions taken in the middle of the solar cycle, at solar maximum, and at solar minimum. We see that, again, both the linear and MLP techniques are able to reproduce the SDO-measured values of $f_{spot}$, $f_{plage}$, and $f_{ntwk}$ at all points in the stellar activity cycle on these timescales.

In Table \ref{tab:corrcoeff}, we list the Pearson correlation coefficients between the HMI derived filling factors and our estimates from the linear and MLP techniques. (For the sake of consistency, note that for both the linear and MLP estimates, we compute correlation coefficients only for results generated using the fraction of data reserved for testing the MLP.) We see that both techniques reproduce the information contained in the HMI filling factors, with the MLP performing slightly better than the linear model on all three filling factors. This may indicate that there is some additional information about the filling factors present in the TSI and S index observations that is not being used by the linear technique. However, given the high degree of correlation produced by both techniques, we can use both estimates of the magnetic filling factors to reduce the effects of activity on observed RVs.

\begin{table}
\centering
\begin{tabular}{c|c|c|c}
     & Spots & Plage & Network  \\
     \hline
    Linear & 0.81 & 0.87 & 0.81 \\
    MLP & 0.85 & 0.87 & 0.89
\end{tabular}
\caption{Pearson correlation coefficients between HMI ground-truth filling factors and the linear and MLP estimates for each class of filling factor.}
\label{tab:corrcoeff}
\end{table}

\subsection{Solar Radial Velocities}

We then fit the HARPS-N solar telescope RVs to Eq. \ref{eq:RVmodel} using the directly-measured SDO filling factors. The results of each fit is given in Table \ref{tab:RVfitParam}. Note that since the linear estimates of $f_{bright}$, $f_{ntwk}$, and $f_{plage}$ are linear transformations of the S index, as shown in Eqs. \ref{eq:LinearBright}, \ref{eq:LinearNtwk}, and \ref{eq:LinearPlage}, we require $D = 0$ to avoid degeneracies when using the filling factor estimates derived from the linear technique. Fitting the linear estimated filling factors to Eq. \ref{eq:RVmodel} without this constraint is equivalent to fitting to $RV = A f_{spot} \left(\frac{df_{spot}}{dt}\right) + B f_{spot} + C' S_{HK} + D' S_{HK} + E'$. We therefore set $D=0$ to avoid having degeneracies between $C$ and $D$ in our fit. No such constraint on $D$ is necessary when considering the SDO or MLP filling factors.

\begin{table*}
\begin{center}
\begin{tabular}{c|c|c|c|c|c}
 Filling Factor Source & $A$ ($10^{5}$ m) & $B$ (m~s$^{-1}$) & $C$ (m~s$^{-1}$) & $D$ (m~s$^{-1}$) & $RV_0$ (m~s$^{-1}$) \\
 \hline
 SDO & $7.7 \pm 1.2$ & $165 \pm 76$ & $244 \pm 15$ & $52 \pm 27$ & $-2.0 \pm 0.3$ \\ 
 Linear & $7.1 \pm 1.3$ & $470 \pm 80$ & $281 \pm 11$ & 0$^*$ & $-1.32 \pm 0.08$ \\ 
 MLP & $7.5 \pm 1.4$ & $479 \pm 218$ & $242 \pm 52$ & $2 \pm 90$ & $-1.3 \pm 1.2$ \\

\end{tabular}
\caption{Best fit coefficients for fitting Eq. \ref{eq:RVmodel} to the HARPS-N solar RVs using SDO filling factors, linear estimates of the filling factors, and MLP estimates of the filling factors. Note that the estimates of $f_{plage}$ and $f_{ntwk}$ derived from the linear technique are both linear transformations of the S index. We therefore require $D = 0$ to avoid degeneracies when using the filling factor estimates derived from the linear technique.}
\label{tab:RVfitParam}
\end{center}
\end{table*}

To ensure our fit is indeed reproducing the suppression of convective blueshift and the photometric RV shift, as expected, we compare the relevant terms of Eq. \ref{eq:RVmodel} to the SDO/HMI estimates of these RV perturbations, as calculated in MH19. In Tables \ref{tab:vcon_corr} and \ref{tab:vphot_corr}, we compare the estimates of $\Delta v_{conv}$ and $\Delta v_{phot}$ computed from Eqs. \ref{eq:vcon} and \ref{eq:vphot} using the filling factors measured by SDO, and estimated using the linear and MLP techniques to the values of $\Delta v_{conv}$ and $\Delta v_{phot}$ derived from HMI observations in \cite{Haywood_2016} and MH19. We see that all of the estimated values of $\Delta v_{conv}$ are highly correlated with the HMI-derived velocities. Our estimates of $\Delta v_{phot}$ are less correlated with the actual photometric shift, but still show good agreement. Interestingly, including the contributions of plage and network regions in Eq.~\ref{eq:vphot} --- that is, adding terms $\propto f_{plage} \times \left(\frac{df_{plage}}{dt}\right)$ and $\propto f_{ntwk} \times \left(\frac{df_{ntwk}}{dt}\right)$ --- does not appear to increase the correlation coefficient. However, we may still conclude that the RVs calculated using Eq.~\ref{eq:RVmodel} indeed do correspond to the combination of suppression of convective blueshift and photometric RV shift described by Eq.~\ref{eq:RVmodel_MH2019}.

\section{Discussion}

\subsection{Filling Factor Estimates}

As shown in Figs. \ref{fig:Results} and \ref{fig:ResultsZoom}, the linear and MLP estimated filling factors successfully reproduce the expected SDO spot, plage, and network filling factors. However, we note that there is a systematic $\sim0.004$ offset between the linear estimates of $f_{ntwk}$ and the SDO measured values. This is likely the result of the significant covariance between $b_1$ and $\mathcal{A} \sigma T_{quiet}^4$ in Eq. \ref{eq:TSIfitstep1}. Any systematic errors in the measured values of $R_{\odot}$ and $T_{quiet}^4$ will change the resulting value of $b_1$, resulting in an offset in the estimated values of $f_{ntwk}$. Small changes to these parameters can dramatically change the observed offset in $f_{ntwk}$: artificially increasing $T_{quiet}^4$ by 0.15 K eliminates the offset entirely. This is well below the precision achieved for measurement of stellar temperatures. While we attain good precision in the solar case, in general linear estimates of $f_{ntwk}$ should assumed to be true up to a constant offset. As stated previously, using these filling factors to remove activity-driven signals from RV measurements only requires values correlated with the filling factor value, making this offset unimportant.

We also note that, while $R_{\odot}$ and $T_{quiet}^4$ are assumed to be constants in our model, they do change in time as the result of physical processes not included in our model. These quantities also vary with wavelength: Since here we are using the Ca II H\&K lines and integrated visible intensity to reproduce filling factors measured at 6173.3 \AA, uncertainties in these parameters associated with their wavelength dependence are inevitable. Indeed, \cite{Meunier_et_al_2010} note that measured filling factors will vary by 20\% to 50\% as a result of these dependencies and other definitional differences: our estimated $f_{ntwk}$ values are certainly consistent with the SDO measured values within these margins.

Fitting the SDO and estimated filling factors to the HARPS-N solar RVs using Eq. \ref{eq:RVmodel} successfully reproduces the expected activity-driven RV variation. As shown in Table \ref{tab:RVfitParam}, for both the SDO measured and MLP-derived filling factors, we see $C >D$. This is consistent with the idea that the denser magnetic interconnections available in photospheric plages are more successful in inhibiting convection, and thus convective blueshifts, than the sparser network magnetizations, as suggested in MH19. Indeed, we see that, using MLP estimates, the network contribution is consistent with zero, and using SDO observations, the network contribution is only $\sim 2 \sigma$ above zero.

The $B$ coefficients, which describe the spot contributions to the suppression of convective blueshift, vary depending on the filling factors used. The linear and MLP estimates of $f_{spot}$ receive a heavy weighting, while the SDO weighting is about a factor of 3 smaller. The MLP estimates also have a contribution only $\sim 2 \sigma$ above zero. However, the Sun is a plage-dominated star, and $f_{spot}$ is about a factor of 100 times smaller than $f_{plage}$, as shown in Fig. \ref{fig:AllVars}. So, while the precise weighting of $f_{spot}$ varies based on the values used, in all cases their contribution to the suppression of convective blueshift will be negligible compared to that of $f_{plage}$. We may therefore conclude that, as suggested by MH19, plage regions are the dominate contribution to the solar suppression of convective blueshift, while spots are the dominant contribution to the photometric RV shift: knowledge of the plage and spot filling factors are therefore sufficient to reproduce $\Delta v_{conv}$ and $\Delta v_{phot}$ respectively.

\subsection{Direct MLP Modelling of Solar RVs}
\label{sec:RVapp}

To see if it is possible for a more refined technique to extract further RV information from our inputs, we fit an MLP directly to the solar RVs using the S-index and TSI as inputs. This is similar to the technique proposed by \cite{debeurs2020}, but replacing the residual cross-correlation function with the S-index and TSI. The hyperparameters of this MLP are the same as those given in Table \ref{tab:NNparam1}. As before, we divide our data into training and test sets, and the quoted residuals are derived from the test set. This fit results in an RMS residual of 0.96 m~s$^{-1}$, (as shown in Table \ref{tab:RVrms}) indicating that there is indeed more RV information to be gained from this set of observations. Interestingly, however, while this RMS value is below the residuals obtained from both sets of estimated filling factors, it is greater than the 0.91 m~s$^{-1}$ residuals obtained by using the direct SDO measurements of the filling factors. This appears to indicate that, while the S index and TSI contain more information than our linear and MLP estimates could obtain, they do not contain \textit{all} the information about the solar plage, spot, and network coverage.

This is unsurprising: we note that network regions can form from decaying plage regions. Due to the geometry of the magnetic flux tubes associated with these regions, a network region may rotate onto the limb, become a plage region as it rotates onto disk center, and then become a network region again as it rotates back onto the limb. The linear technique directly uses the different temperature contrasts of network and plage to provide a useful first pass at differentiating these regions, but does not capture these links between them. That is, there are additional physical effects that further complicate the relationship between photometry, spectroscopy, filling factors, and RVs \citep{Miklos_2020}. While the underlying behavior of the MLP is unknown, it likely employs a similar, slightly more complex technique to differentiate the two classes of regions. The magnetic intensification effect, which strengthens lines in the presence of a magnetic field (e.g., \citealt{Leroy1962, StifftLeone2003}), has an RV signal which depends on the overall filling factor, as well as a given line's wavelength, effective Land\'e $g$ value, and the magnetic field strength \citep{Reinersetal2013}. HMI monitors the photospheric 6173.3 \AA\ iron line: these wavelength-dependent effects mean that the filling factors derived from HMI may not be consistent with those derived from the chromospheric calcium H and K lines. The center-to-limb dependence of the calcium H and K lines are different than the 6173.3 \AA\ line as well, which could lead to mismatches in the derived filling factors as a function of rotational phase. More complicated linear and MLP-based filling factor estimates could use spectroscopic measurements of additional absorption lines, and photometric measurements integrated over different wavelength bands to compensate for these effects, and to exploit different wavelength-dependent contrasts of each feature to better separate these three classes of magnetic active regions.

\begin{figure} 
\begin{center} 
\includegraphics[width=.5\textwidth]{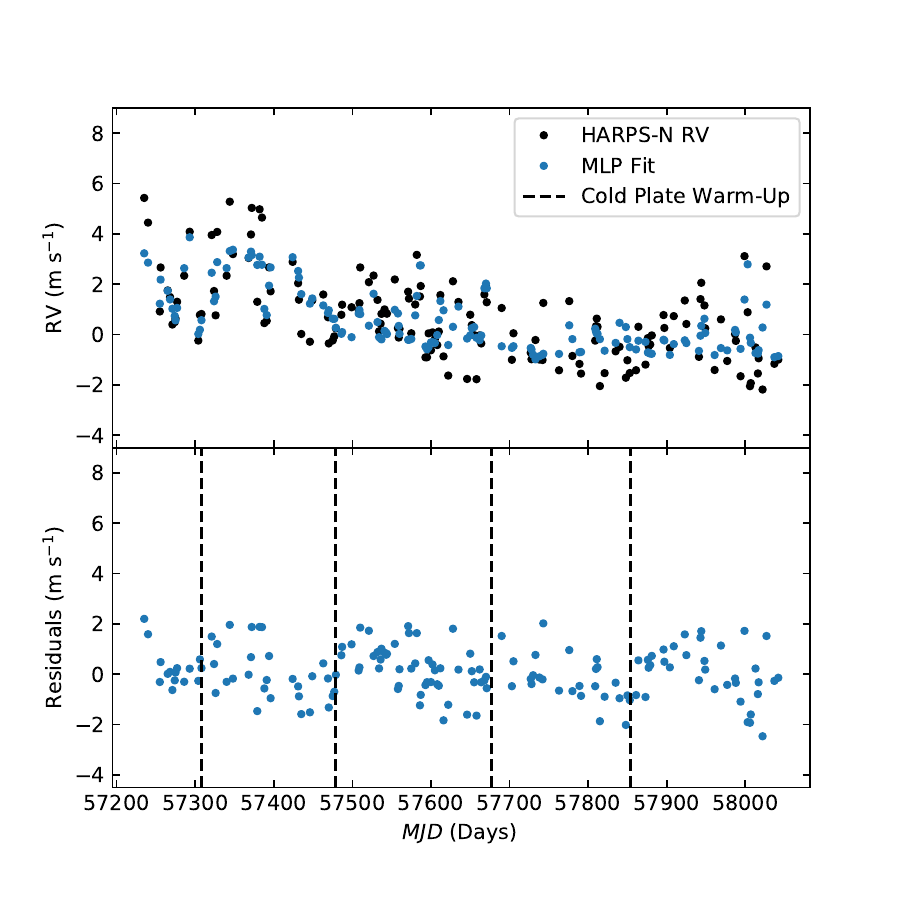} 
\caption{\label{fig:MLPfit} MLP fit to the HARPS-N solar telescope data. HARPS-N RVs are shown in black, and MLP estimates of the RVs are in blue. Fit residuals are shown in the bottom panels: HARPS-N cryostat warm-up dates (\emph{see text}) are indicated with black dashed lines.
}
\end{center} 
\end{figure}

The direct MLP fit to the solar RVs and its residuals are plotted in Fig.~\ref{fig:MLPfit}. The effects of HARPS-N cryostat cold plate warm-ups, discussed in \cite{ACC2019, dumusque2020years}, are clearly visible in the fit residuals, indicating that the MLP is not "learning" instrumental systematics, and that the residual RV variations below this level are likely dominated by a combination of instrumental systematics and activity processes not reflected by variations in the S-index and TSI. Further work is necessary to identify these remaining activity processes, and to disentangle them from instrumental effects.

\subsection{Application to the Stellar Case}
\label{sec:StellarOutlook}
The techniques developed in this work should be applicable to Sun-like stars with the proper observational cadence. To reproduce properly scaled filling factors, the linear technique requires precise knowledge of radius, effective temperature, and distance to the target, as well as the temperature contrasts of the plage, spots, and network. The effective temperature may be calculated spectroscopically \citep{SPC2012}, while the temperature contrasts may be assumed to be Sun-like in the case of G-class stars. The stellar radius may then be calculated photometrically, using the spectroscopic temperature as a prior. The stellar distance may be straightforwardly determined through parallax measurements, while the rotation period may be obtained via photometry or through RV measurements.

Although the techniques presented here assume a plage-dominated star, it is straightforward to rework Eqs.~\ref{eq:TSIfitstep1} and \ref{eq:TSIfitstep2} for a spot-dominated target: in this case, the S index is assumed to be correlated with spot-driven variations of the TSI, with positive deviations indicating the presence of bright, plage regions.

While properly scaling the linear estimate of $f_{plage}$ requires observations near stellar minimum, modelling RV variations only requires time series which are proportional to $f_{spot}$, $f_{plage}$ and $f_{ntwk}$---in this case, this offset is unimportant, and no additional constraint is placed on the stellar observations. Furthermore, the values of $\Delta T_{spot}$, $\Delta T_{plage}$ and $\Delta T_{ntwk}$ may be absorbed into the fit coefficients in Eqs.~\ref{eq:TSIfitstep1}, \ref{eq:TSIfitstep2}, and \ref{eq:RVmodel}, further simplifying matters.

The MLP machine learning technique, in contrast, requires less knowledge of the target star: while the mathematical and physical transformations learned by MLP are unknown to the user, the MLP is presumably learning a more sophisticated version of the linear technique, and implicitly "learns" the solar values for feature temperature contrast and quiet temperature as it identifies higher-order correlations between the TSI, S index, and filling factors. This makes the MLP straightforward to implement when precise contrast values are unknown. Furthermore, since the MLP uses no timing information, it places no constraints on the observational cadence or baseline: it only requires simultaneous photometric and spectroscopic measurements.

However, since ground-truth filling factors can only be directly measured in the solar case, the MLP must be trained using solar data. Its stellar application is therefore limited to Sun-like stars (that is, stars with very similar surface filling factors as the Sun, or possibly even only solar twins), making it less generalizable to other stellar targets. Since stars other than the Sun cannot be resolved spatially at high resolution, assessing just how "Sun-like" a target needs to be for the machine learning technique to yield meaningful results is challenging. One possibility is to generate synthetic stellar images for targets with a variety of spectral types, activity levels, feature contrasts, and viewing angles using SOAP 2.0 \citep{dumusque2014}, StarSim \citep{StarSim}, or a similar platform, computing the light curve and S-index for these images, and seeing if an MLP trained on the solar case reproduces the filling factors expected for each image. Such an analysis is beyond the scope of this work.

\section{Conclusions}

We assess two techniques to extract spot, plage, and network filling factors using simultaneous spectroscopy and photometry. The first technique involves a straightforward analytical manipulation of the S-index and TSI time series, while the second uses a neural network machine learning technique known as a Multilayer Perceptron (MLP) trained on ground-truth filling factors derived from full-disk solar images. Both techniques yield filling factor estimates which are highly correlated with values derived from full-disk solar images, with Spearman correlation coefficients ranging from 0.81 and 0.89 from each technique.

We show that decorrelating a nearly-three-year time series of solar RVs using HMI-observed spot, plage, and network filling factors effectively reproduces the expected RV variations due to the convective blueshift and rotational imbalance due to flux inhomogeneities, reducing the residual activity-driven RVs more than the typical technique of decorrelating using spectroscopic activity indices alone. Fitting to HMI filling factors reduces the RV RMS from 1.64 m~s$^{-1}$ to 0.91 m~s$^{-1}$, while fitting to the S-index alone results in an RMS variation of 1.10 m~s$^{-1}$. Including this additional information about spots, plage, and network thus accounts for an additional $\sqrt{(1.10 \mathrm{m~s}^{-1})^2 -  (0.91 \mathrm{m~s}^{-1})^2} = 0.62 \mathrm{m~s}^{-1}$ of RMS variation. The filling factor estimates from both the linear and MLP techniques offer some improvement to the RMS residuals beyond what is obtained from only the S-index. Decorrelating with the linear estimates reduces the RMS variation to 1.04 m~$s^{-1}$, and the MLP estimated filling factors reduces the RMS to 1.02 m~s$^{-1}$.

Using a MLP trained directly on the solar RVs, we reduced the RMS to 0.96 m~s$^{-1}$. While this indicates that the S-index and TSI contain more RV information than obtained by either estimate of our filling factors, it does not lower the RMS RVs below the 0.91 m~s$^{-1}$ limit obtained using direct measurements of the magnetic filling factors. This suggests that, while our initial estimates of $f_{spot}$, $f_{plage}$, and $f_{ntwk}$ are highly correlated with the expected value, more information is needed to fully characterize these feature-specific filling factors. To match the performance of the HMI filling factors, a more sophisticated version of this technique, using additional spectral lines and photometric bands will likely be necessary.

Both the analytical and machine learning techniques may be used to extract filling factors on other stars: the analytical technique is more widely generalizable, but requires detailed knowledge of the star and good temporal sampling, ideally with observations of the target at activity minimum. The machine learning technique, in contrast, requires no additional knowledge of the target star, and applies no constraints on the observing schedule---however, it is only applicable to stars with very similar filling factor properties as the Sun.

\acknowledgments

This work was primarily supported by NASA award number NNX16AD42G and the Smithsonian Institution. The solar telescope used in these observations was built and maintained with support from the Smithsonian Astrophysical Observatory, the Harvard Origins of Life Initiative, and the TNG.

This material is also based upon work supported by NASA under grants No. NNX15AC90G and NNX17AB59G issued through the Exoplanets Research Program. The research leading to these results has received funding from the European Union Seventh Framework Programme (FP7/2007-2013) under grant Agreement No. 313014 (ETAEARTH). 

The HARPS-N project has been funded by the Prodex Program of the Swiss Space Office (SSO), the Harvard University Origins of Life Initiative (HUOLI), the Scottish Universities Physics Alliance (SUPA), the University of Geneva, the Smithsonian Astrophysical Observatory (SAO), and the Italian National Astrophysical Institute (INAF), the University of St Andrews, Queen's University Belfast, and the University of Edinburgh.

We would like to acknowledge the excellent discussions and scientific input by members of the International Team, "Towards Earth-like Alien Worlds: Know thy star, know thy planet", supported by the International Space Science Institute (ISSI, Bern).

This work was partially performed under contract with the California Institute of Technology (Caltech)/Jet Propulsion Laboratory (JPL) funded by NASA through the Sagan Fellowship Program executed by the NASA Exoplanet Science Institute (R.D.H.).

S.H.S. is grateful for support from NASA Heliophysics LWS grant NNX16AB79G and NASA XRP grant 80NSSC21K0607..

A.M. acknowledges support from the senior Kavli Institute Fellowships.

A.C.C. acknowledges support from the Science and Technology Facilities Council (STFC) consolidated grant number ST/R000824/1.

X.D. is grateful to the Branco-Weiss Fellowship for continuous support. This project has received funding from the European Research Council (ERC) under the European Unions' Horizon 2020 research and innovation program (grant agreement No. 851555).

C.A.W. acknowledges support from Science and Technology Facilities Council grant ST/P000312/1.

H.M.C. acknowledges financial support from the National Centre for Competence in Research (NCCR) PlanetS, supported by the Swiss National Science Foundation (SNSF), as well as UK Research and Innovation Future Leaders Fellowship grant No. MR/S035214/1

We thank the entire TNG staff for their continued support of the solar telescope project at HARPS-N.

%%% FACILITIES AND SOFTWARE %%%
\facilities{TNG: HARPS-N \citep{Cosentino2014}, Solar Telescope \citep{Dumusque:2015, Phillips2016}}
\software{HARPS-N DRS, Python \citep{Python}: \texttt{NumPy} \citep{NumPy}, \texttt{SciPy} \citep{SciPy}, \texttt{scikit-learn} \citep{scikit-learn}}

%%% REFERENCES %%%
\bibliographystyle{aasjournal}
\bibliography{references}

\end{document}